\documentclass{jps-cp}
\usepackage{txfonts} 
\usepackage{pdfpages}
\usepackage{subfigure}
\usepackage{lineno}
\usepackage{xfrac}

\title{~\\ Transverse Spin Transfer of $\Lambda$ and $\overline{\Lambda}$ Hyperons in Polarized $p$+$p$ Collisions at $\sqrt{s} = 200$ GeV at RHIC-STAR
}

\author{Yike Xu$^{1}$, for STAR Collaboration}
\inst{{\bf 1} Institute of Frontier and Interdisciplinary Science \& Key Laboratory of Particle Physics and Particle Irradiation (MOE), 
Shandong University, Qingdao, Shandong, 266237, China}

\email{yxu@rcf.rhic.bnl.gov}

\abst{The transverse spin transfer, $D_{TT}$, of $\Lambda$ and $\overline{\Lambda}$ hyperons in $p$+$p$ collisions is expected to be sensitive to the $s$ and $\bar{s}$ quark transversity distributions in the proton and to the transversely polarized fragmentation functions. The STAR experiment has published the first measurement of the transverse spin transfer of $\Lambda$ and $\overline{\Lambda}$ hyperons in transversely polarized $p$+$p$ collisions at $\sqrt{s}$ = 200 GeV within pseudo-rapidity $|\eta| < 1.2$ and for the transverse momenta up to 8 GeV/$c$ based on the data taken in 2012. In 2015, a data sample of $p$+$p$ collisions at $\sqrt{s}$ = 200 GeV, about two times larger than the 2012 data, was collected. This contribution presents the preliminary results of the transverse spin transfer, $D_{TT}$, of $\Lambda$ and $\overline{\Lambda}$ hyperon versus transverse momentum and fractional momentum of the hyperon within a jet, based on 2015 data.}

\begin{document}
\maketitle

\section{Introduction}
\label{sec:intro}
The $\Lambda$ hyperon contains a strange constitute quark, which is expected to carry most of $\Lambda$ spin. The  $\Lambda$ polarization, $P_{\Lambda(\bar{\Lambda})}$, can be determined from the angular distribution of its weakly decayed daughters. The transverse spin transfer to $\Lambda$ hyperons in hadron-hadron collisions is thus expected to be able to provide insights into the strange quark transversity and polarized fragmentation functions~\cite{Qinghua2004,Qinghua2006dtt}.
The transversity distribution remains largely unknown due to its chiral-odd nature~\cite{chiral-odd1,chiral-odd2}.


The transverse spin transfer, $D_{TT}$, of $\Lambda$ hyperons in $p$+$p$ collisions is defined as:

\begin{gather}
D_{TT}^{\Lambda} \equiv \frac{d\sigma^{(p^{\uparrow}p \to \Lambda^{\uparrow}X)} - d\sigma^{(p^{\uparrow}p \to \Lambda^{\downarrow}X)}}{d\sigma^{(p^{\uparrow}p \to \Lambda^{\uparrow}X)} + d\sigma^{(p^{\uparrow}p \to \Lambda^{\downarrow}X)}} = \frac{d\delta\sigma^{\Lambda}}{d\sigma^{\Lambda}},  
\label{defination}
\end{gather}
where $p^\uparrow$/$p^\downarrow$ and $\Lambda^\uparrow$/$\Lambda^\downarrow$ denote the positive/negative transverse polarizations of the colliding proton and the $\Lambda$ hyperon. $d\delta\sigma^{\Lambda}$ is the transversely polarized cross section and $d\sigma^{\Lambda}$ is the unpolarized cross section. Under the factorization framework, the polarized cross section can be written as the convolution of parton transversity distribution, polarized partonic cross section and the polarized fragmentation function.

STAR has published the first measurement of $D_{TT}$ versus transverse momentum ($p_T$) of $\Lambda$ hyperons in $p$+$p$ collisions at $\sqrt{s} = 200$ GeV with data collected in 2012~\cite{AdamJimprovedDTTMeasurement}.
This contribution presents the new preliminary results of $D_{TT}$ for $\Lambda(\bar{\Lambda})$ in $p$+$p$ collisions at $\sqrt{s} = 200$ GeV, using data taken at STAR in 2015 with about twice the statistics of the 2012 dataset.
We present $D_{TT}$ as a function of hyperon transverse momentum $p_T$, as well as the first measurement of $D_{TT}$ versus fractional momentum of the hyperon within a jet, which provides direct information on the polarized fragmentation function.

\section{Dataset and Hyperon Reconstruction}

This analysis is based on the $p$+$p$ collision data at 200 GeV taken in 2015 with an average beam polarization of $56\%$ and an integrated luminosity of $52$ $\rm{pb^{-1}}$.
For this analysis, several STAR sub-detectors are used, including the Time Projection Chamber (TPC), ElectroMagnetic Calorimeter (EMC), and Time Of Flight (TOF) detector.
Hard scattering events are selected with a Jet Patch trigger which is based on energy deposits in the EMC. 

The $\Lambda(\bar{\Lambda})$ candidates are identified from the weak decay channels, $\Lambda\to p\pi^-$ ($\bar{\Lambda} \to \bar{p}\pi^+$). Pairs of proton and pion candidate tracks measured in the TPC are used to reconstruct the $\Lambda(\bar{\Lambda})$ candidates. TOF information is also used to improve particle identification. Then, a series of topological cuts are tuned to further reduce the background. 

\begin{figure}[h]
    \centering
    \subfigure[$\Lambda$]{
    \includegraphics[width=0.45\columnwidth]{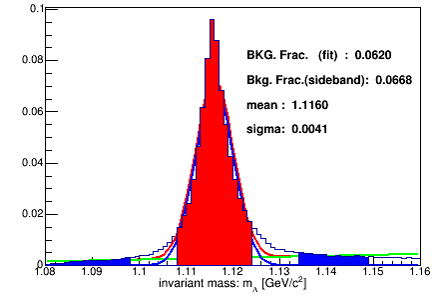}}
    \centering
    \subfigure[$\bar{\Lambda}$]{
    \includegraphics[width=0.45\columnwidth]{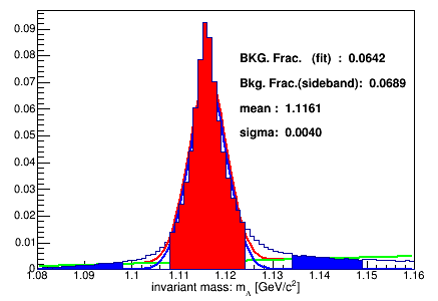}}
    \caption{The invariant mass distributions for $\Lambda$ (a) and $\bar{\Lambda}$ (b) candidates within 3 $<$ $p_T$ $<$ 4 GeV/$c$ after applying the selection cuts. The side-band regions are shown in blue and the signal regions are in red.}
    \label{fig:lambda}
\end{figure}

The spin transfer of the $\Lambda$ hyperon is obtained by subtracting the contribution from background, $D_{TT} = (D_{TT}^{raw}-rD_{TT}^{bkg})/(1-r)$, where $D_{TT}^{raw}$ and $D_{TT}^{bkg}$ are the spin transfers for signal and side-band regions, and $r$ is the background fraction. The side-band method is used as in~\cite{AdamJimprovedDTTMeasurement} to estimate the background fraction, which is less than 10\%. Figure~\ref{fig:lambda} shows examples of $\Lambda$ and $\bar{\Lambda}$ candidates invariant mass distribution.

\section{$D_{TT}$ extraction method}

We followed the method used in the 2012 dataset analysis~\cite{AdamJimprovedDTTMeasurement} to extract the $D_{TT}$.
For the transverse spin transfer $D_{TT}$ measurements, the hyperon polarization direction is defined as the transverse polarization direction of the outgoing parton, which can be obtained by rotating the polarization vector of the incoming parton along the normal direction of the partonic scattering plane~\cite{collins,AdamJimprovedDTTMeasurement}. The scattering plane is spanned by the momentum directions of incoming and outgoing partons. 
The reconstructed jet axis is used as the surrogate of the momentum direction of the fragmenting parton~\cite{AdamJimprovedDTTMeasurement}. The anti-$k_T$ algorithm is used to reconstruct jets, similar as in Ref.~\cite{jetrec1,STAR:2019yqm}, with the resolution parameter of $R$ = 0.6. We require jet transverse momentum $p_T>5$ $\rm{GeV}$/$c$ and $-0.7<\eta_{det}<0.9$. 
The detector pseudorapidity, $\eta_{det}$, is defined as relative to the center of the STAR detector. 
The correlation between $\Lambda(\bar{\Lambda})$ candidates and the reconstructed jets is made by constraining the distance $\Delta R=\sqrt{(\Delta\eta)^2+(\Delta\phi)^2 }$ between $\Lambda(\bar{\Lambda})$ candidates and the reconstructed jets in $\eta-\phi$ space. The $\Lambda(\bar{\Lambda})$ hyperons in the near-side of jets $(\Delta R<0.6)$ are kept for the $D_{TT}$ analysis.

$D_{TT}$ is measured from a cross-ratio asymmetry using $\Lambda$ counts for opposite beam polarizations within small $\cos \theta^*$ bins~\cite{AdamJimprovedDTTMeasurement}:
\begin{equation} 
D_{TT} = \frac{1}{\alpha P_{beam} \left \langle \cos\theta^* \right \rangle } \frac{\sqrt{N^{\uparrow}(\cos \theta^*)N^{\downarrow}(-\cos \theta^*)}-\sqrt{N^{\uparrow}(-\cos\theta^*)N^{\downarrow}(\cos\theta^*)}}{\sqrt{N^{\uparrow}(\cos\theta^*)N^{\downarrow}(-\cos\theta^*)}+\sqrt{N^{\uparrow}(-\cos\theta^*)N^{\downarrow}(\cos\theta^*)}}.
\end{equation}
where the $\theta^*$ is the angle between the $\Lambda(\bar{\Lambda})$ polarization direction and the (anti-) proton momentum in the $\Lambda(\bar{\Lambda})$ rest frame, the $\left \langle \cos\theta^* \right \rangle$ is the average value of each $\cos\theta^*$ bin, $N^{\uparrow/\downarrow}$ is the $\Lambda$ count in a $\cos\theta^*$ bin when the polarization of the beam is upward or downward, $\alpha=0.732\pm0.014$~\cite{PDG} is the decay parameter and $P_{beam}$ is the beam polarization.
With the cross-ratio method, the relative luminosity and the detector acceptance are both cancelled, which helps to reduce systematic uncertainties.

\section{ $D_{TT}$ Result}

In this section we present $D_{TT}$ as a function of hyperon $p_T$ and momentum fraction $z$.
\subsection{$D_{TT}$ versus $p_T$}

\begin{figure}[h]
    \centering
    \includegraphics[width=0.75\textwidth]{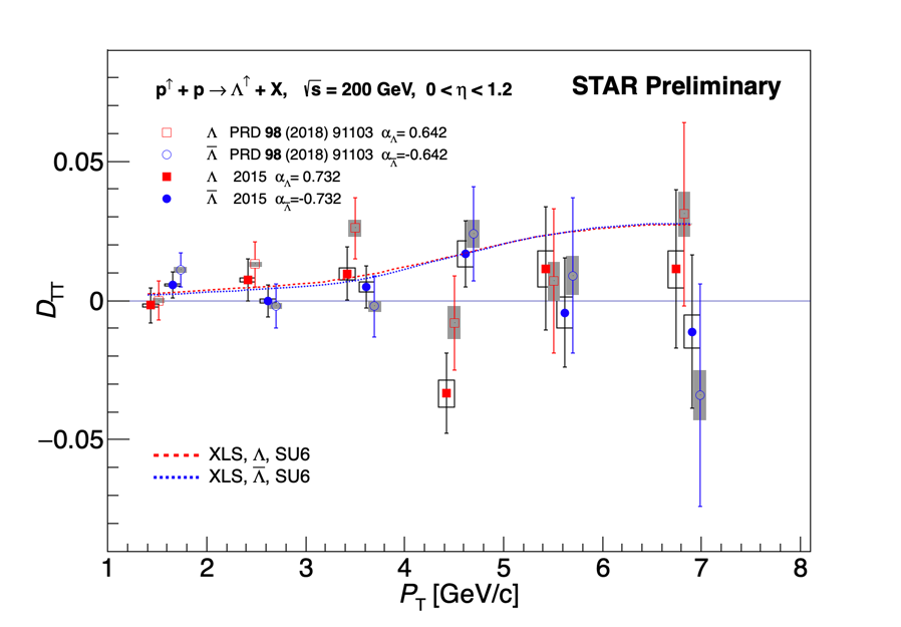}
    \caption{Preliminary results of $D_{TT}$ for $\Lambda$ and $\bar{\Lambda}$ from STAR 2015 data versus hyperon $p_T$, together with previously published results~\cite{AdamJimprovedDTTMeasurement}.The theory calculations are presented in dotted line for comparison.
The $\bar{\Lambda}$ results and the previously published results have been shifted slightly to larger $p_T$ values for clarity. }
    \label{fig:figure_DTT_vs_lambdapt}
\end{figure}

Figure~\ref{fig:figure_DTT_vs_lambdapt} shows the new preliminary $D_{TT}$ results from 2015 data, together with previously published results from the 2012 dataset~\cite{AdamJimprovedDTTMeasurement}, versus $\Lambda(\bar{\Lambda})$ $p_T$ in the positive pseudo-rapidity region relative to the polarized beam. The statistical uncertainties of the new results are improved by a factor of $1/\sqrt 2$ compared to previous results. The new results are consistent with previously published results, and are also consistent with model predictions~\cite{Qinghua2006dtt}. 

The systematic uncertainties include contributions from hyperon decay parameter, background fraction estimation, beam polarization measurement and trigger bias. The trigger bias is the dominant source of the systematic uncertainties, which is estimated by comparing the $D_{TT}$ values obtained from a theoretical model before and after applying the trigger conditions on a Monte-Carlo data sample produced with Pythia6.4 event generator~\cite{Pythia} and STAR detector simulation by Geant~\cite{geant}. 


\subsection{$D_{TT}$ versus $z$}

\begin{figure}[h]
    \centering
    \includegraphics[width=0.65\textwidth]{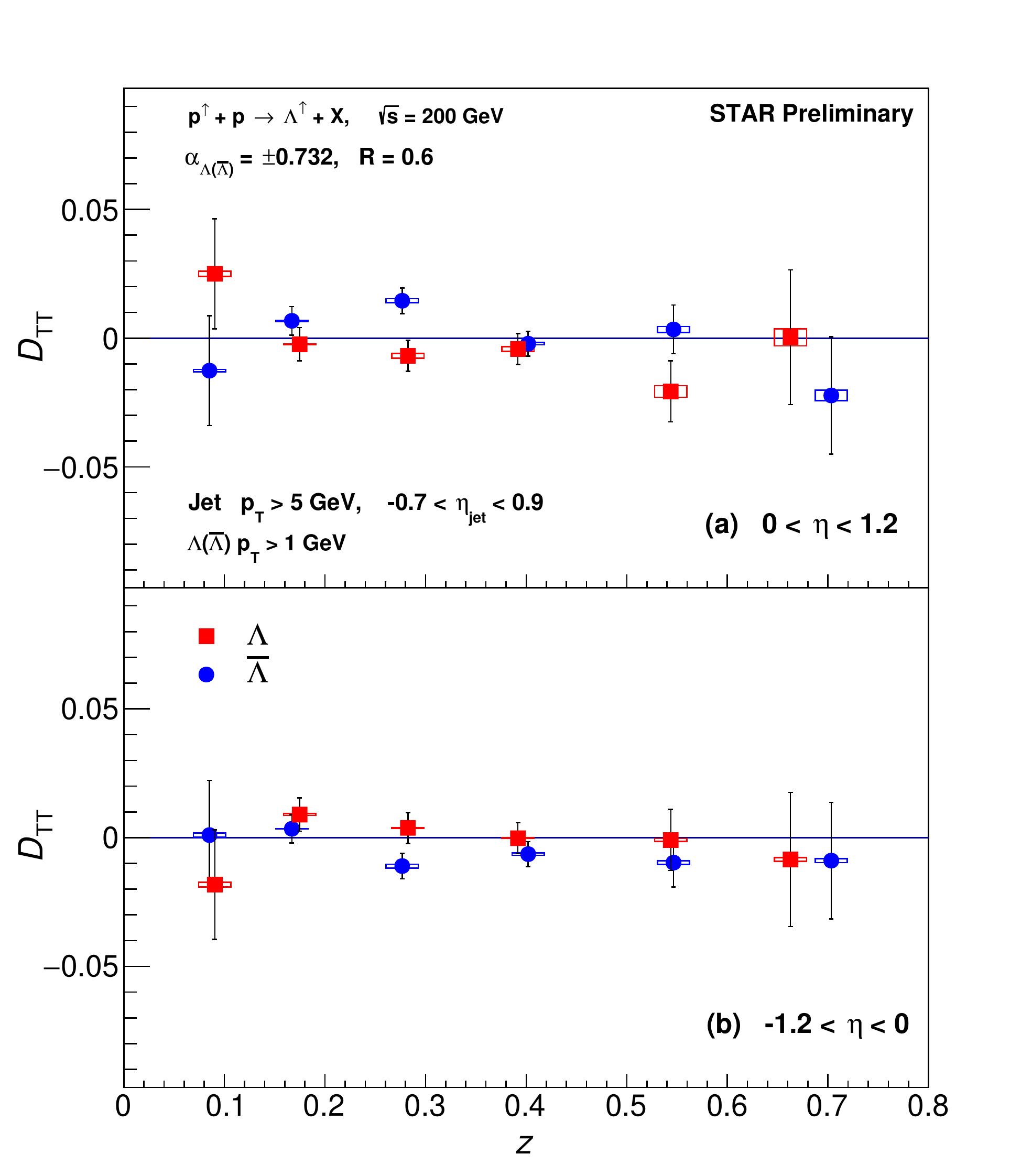}
    \caption{Preliminary results of transverse spin transfer $D_{TT}$ versus $z$, for $\Lambda$ and $\bar\Lambda$ in $p$+$p$ collisions at $\sqrt{s}=200$ GeV at STAR.  }
    \label{fig:figure_DTT_vs_lambdaz}
\end{figure}

$D_{TT}$ versus the hyperon fractional momentum within a jet can provide direct information about the transversely polarized fragmentation function. We perform the first measurement of $D_{TT}$ versus $z$ in $p$+$p$ collisions with $z$ defined as: $z \equiv \overrightarrow{p}_{\Lambda} \cdot \overrightarrow{p}_{jet} / \mid \overrightarrow{p}_{jet} \mid^2$, where $\overrightarrow{p}_{jet}$ is measured from the reconstructed jet with TPC tracks and EMC energy deposits. The $z$ value obtained with reconstructed jets is the detector level $z$. In theoretical calculations, all produced particles are included in the jet, corresponding to the particle level $z$. In this analysis, we first measure $D_{TT}$ in different detector $z$ bins. Then, the Monte-Carlo data sample based on Pythia6 and Geant is used for correcting the detector level $z$ to the particle level $z$.
Figure~\ref{fig:figure_DTT_vs_lambdaz} shows $D_{TT}$ versus $z$ within a jet in $p$+$p$ collisions at 200 GeV. The results are consistent with zero within uncertainties. 

\section{Summary and Outlook}
Transverse spin transfer, $D_{TT}$, in $p$+$p$ collisions can provide access to the strange quark transversity distributions in the proton and the transversely polarized fragmentation functions.
New preliminary results of $D_{TT}$ with the dependence on hyperon $p_T$ in $p$+$p$ collisions at 200 GeV from STAR 2015 dataset are reported, with twice the statistics of the previously published results. The new results are consistent with previous measurements, and are consistent with zero within uncertainties. The first measurement of $D_{TT}$ versus hyperon's fractional momentum $z$ within a jet through the same dataset is also reported, which directly probes the transversely polarized fragmentation function. 

STAR has expanded its acceptance by installing a series of detector upgrades in the forward rapidity region ($2.5<\eta<4$). More $p$+$p$ collision data will be collected at STAR in 2022 at 510 GeV and in 2024 at 200 GeV. Those detector upgrades allow for a rich $\Lambda$ physics program in the forward rapidity region, including not only spin transfer measurements but also the transverse hyperon polarization in unpolarized collisions.


\section*{Acknowledgements}

The author is supported partially by the National Natural Science Foundation of China under No. 12075140.

\end{document}